\documentstyle[amssymb,prb,aps]{revtex}

\draft
\input psfig.tex

\begin{document}
\title{Array of Bose-Einstein condensates under
time-periodic Feshbach-resonance management}

\author{F. Kh. Abdullaev$^{a,b}$, E. N. Tsoy$^a$
\footnote{Corresponding author, e-mail: etsoy@physic.uzsci.net},
B. A. Malomed$^c$, and R. A. Kraenkel$^b$}

\address{$^a$ Physical-Technical Institute
  of the Uzbek Academy of Sciences,\\
2-B, Mavlyanov street, Tashkent, 700084, Uzbekistan\\
$^b$ Instituto de Fisica Teorica, UNESP, R. Pamplona 145,\\
01405-900 Sao Paulo, Brazil\\
$^c$ Department of Interdisciplinary Studies, Faculty of\\
Engineering, Tel Aviv University, Tel-Aviv 69978, Israel}
\date{\today}
\maketitle

\begin{abstract}
The dynamics of a discrete soliton in an array of Bose-Einstein condensates
under the action of a periodically time-modulated atomic scattering length
(``Feshbach-resonance management, FRM'') is investigated. The cases of both
slow and rapid modulation, in comparison with the tunneling frequency, are
considered. We employ a discrete variational approach for the analysis of
the system. The existence of nonlinear resonances and chaos is predicted at
special values of the driving frequency. Soliton splitting is observed in
numerical simulations. In the case of the rapid modulation, we derive an
averaged equation, which is a generalized discrete nonlinear Schr\"odinger
equation, including higher-order effective nonlinearities and intersite
nonlinear interactions. Thus the predicted discrete FRM solitons are a
direct matter-wave analog of recently investigated discrete
diffraction-managed optical solitons. \end{abstract}

\pacs{ 03.75.Lm, 42.65.Wi, 05.45.-a}


\section{Introduction}

Discrete solitons in nonlinear lattices with periodically varying parameters
have recently attracted much attention. System belonging to this type
include arrays of optical waveguides subject to periodic diffraction
management \cite{Atch,Abl1} and waveguide arrays with a periodic modulation
of the tunnel-coupling constant \cite{Pesch}. The corresponding model is
typically based on the discrete nonlinear Schr\"{o}dinger (DNLS) equation,
with the coefficient in front of the second-finite-difference term varying
along the propagation distance (formally, it looks like periodic time
modulation of the coefficient). It was shown that, in the case of rapid and
strong variations of the coupling constant, a stable breathing discrete
soliton can exists \cite{Abl1} (the so-called diffraction-managed soliton).
On the other hand, application of a relatively slow weak or moderate
modulation at a resonant frequency results in a splitting of the discrete
soliton \cite{Pesch}.

In periodically modulated DNLS systems of another type, the coefficient of
the on-site cubic nonlinearity is subject to the modulation. In terms of
nonlinear optics, these may be arrays of waveguides which have a layered
structure, with the strength \cite{Gai}, or even sign \cite{Mal2}, of the
nonlinearity alternating between layers. An alternative, and actually more
straightforward, physical realization of this type of the lattice is offered
by an array of droplets of a Bose-Einstein condensate (BEC) trapped in a
deep optical lattice \cite{Abd3,Tromb}, with the BEC scattering length
oscillating in time. The latter type of the time-modulation may be provided
by ac magnetic field tuned to the Feshbach resonance, as it was predicted
theoretically \cite{Feshbach} and demonstrated experimentally \cite
{FeshbachExperiment}. By analogy with the well-known techniques of the
dispersion management \cite{DM} and the above-mentioned diffraction
management \cite{Atch} in nonlinear optics, this time-modulation technique,
applied to BEC, may be called Feshbach-resonance management (FRM). Very
recently, it has been demonstrated that FRM provides for an effective
mechanism of stabilization of two-dimensional BECs, even in the absence of
the dc-magnetic-field trap \cite{ACKMSU}. One-dimensional solitons subject
to the action of FRM were also recently studied, which reveals stable
breathers oscillating between the Gaussian and Thomas-Fermi shapes, and
stable breathers of other types \cite{FRM}.

The aim of the present work is to consider the dynamics of solitons in the
one-dimensional DNLS model with the nonlinearity subject to periodic
modulation. We will treat the cases of both relatively slow and rapid
modulations. In the former case, we will apply an analytical variational
approximation (VA), which was developed for one-dimensional lattice models
in Ref.~\cite{Malom}, and direct simulations, to study resonances and
splitting in the discrete-soliton dynamics (a recent review of the VA
technique can be found in Ref.~\cite{Progress}). In the latter case, using
the multiscale method \cite{Kath}, we will derive an averaged equation,
which has the form of a generalized DNLS equation with new nonlinear on-site
and inter-site terms. Using this equation and VA, we will analyze the
structure of average discrete-soliton solutions.

\section{The model}

\label{Sect:Gen}

We formulate the model in terms of a BEC trapped in a deep optical lattice,
which is created by the interference of two counterpropagating optical
beams. The dynamics of a BEC is governed by the Gross-Pitaevskii equation
\cite{Dalf}:
\begin{equation}
i\hbar \frac{\partial \psi }{\partial t}=-\frac{\hbar ^{2}}{2m}\Delta \psi
+V({\bf r})\psi +g(t)|\psi |^{2}\psi \ ,  \label{GP}
\end{equation}
where $V({\bf r})=V_{0}(x,y)\sin ^{2}(kz)$ is the optical potential, and
$g(t)=4\pi \hbar ^{2}a_{s}(t)/m$. Here $a_{s}$ is the time-varying atomic
scattering length, and $m$ is the atomic mass. As it was mentioned above,
the time dependence of $a_{s}$ can be induced by ac magnetic field (or laser
radiation with a time-modulated intensity) applied to the condensate. Due to
the periodicity of $V({\bf r})$, for a weakly coupled array of BECs, one can
present a solution as
\begin{equation}
\psi =\sum_{n}u_{n}(t)\Phi ({\bf r}-{\bf r}_{n})\ ,  \label{Expans}
\end{equation}
where the function $\Phi({\bf r}-{\bf r}_{n})$ is assumed to be
strongly localized around $n$-th site. Substituting Eq.~(\ref{Expans}) into
Eq.~(\ref{GP}), integrating over the transverse coordinates, and taking into
account the exchange integrals only for neighboring sites, one arrives at a
DNLS equation with a variable coefficient in front of the nonlinear term
\cite{Abd3,Tromb}:
\begin{equation}
i\dot{u}_{n}+\frac{1}{2}\left( u_{n+1}+u_{n-1}-2u_{n} \right) +a(t)\left|
u_{n}\right|^{2}u_{n}=0\ .  \label{DNLS}
\end{equation}
Here the overdot stands for the time derivative, time is made dimensionless
by means of the rescaling $t\rightarrow t\hbar /K$, where $K$ is the
tunnel-coupling parameter between adjacent wells in the optical lattice
\cite{Abd3,Tromb}, and
\begin{equation}
a(t)=a_{0}+a_{1}\sin \left( \omega t\right)   \label{a(t)}
\end{equation}
is a coefficient proportional to minus the atomic scattering length in the
BEC. Equation (\ref{DNLS}) describes the dynamics restricted within the
lowest Bloch zone. Account of inter-zone transitions requires an
extension of the DNLS model \cite{Alfim}.

Though the BEC system described above is the most relevant physical
realization of Eq.~(\ref{DNLS}), the same model also applies to an array of
periodically modulated optical waveguides, with $t$ being the propagation
distance, rather than time. Without loss of generality, one can set $a_{0}=1$
and $a_{0}=-1$ in Eq.~(\ref{a(t)}) for the cases of the negative and
positive scattering lengths (repulsion and attraction between atoms),
respectively. The wave function $u_{n}(t)$ is normalized so that the
dynamical invariant of Eq.~(\ref{DNLS}),
\begin{equation}
W=\sum_{n=-\infty }^{\infty }|u_{n}|^{2},  \label{norm}
\end{equation}
is the total number of particles. The characteristic length of the
system $2\pi /k \sim 1~\mu {\rm m}$, $V_{0}\approx \hbar
^{2}k^{2}/m$, and the atomic population in each well is $\sim 10^{3}$
atoms. The characteristic frequency for the tunneling between wells is
$\Omega _{L}=K/\hbar \sim 10~{\rm Hz}$, and the separation between the
energy levels in a single well is $\Omega \gtrsim 100~{\rm Hz}$.
Therefore, it makes sense to consider the variation of the driving
frequency $\omega $ in the interval $\Omega_{L}<\omega K/\hbar <\Omega$.

First, we consider stationary pulse-shaped solutions of the unperturbed DNLS
equation with $a_{1}=0$ in Eq.~(\ref{a(t)}) in the form (see, e.g., Ref.~
\cite{Scott,Aceves})
\begin{equation}
u_{n}(t)=Q_{n}\exp (i\kappa n-i\chi t)\ ,  \label{Soliton}
\end{equation}
where $\kappa $ is a wave number and $\chi $ is a frequency. As it follows
from the dispersion relation for the linearized equation~(\ref{DNLS}), a
localized solution with the maximum of $Q_{n}$ centered at some fixed point
exists only for particular values of $\kappa $ at which the group velocity
vanishes, so we take $\kappa =0$ for $a_{0}=1$, or, equivalently, $\kappa
=\pi $ for $a_{0}=-1$.

The fundamental soliton for $a_{0}=1$ was studied in detail as a numerical
solution to the nonlinear eigenvalue problem with zero boundary conditions
\cite{Scott,Aceves,Cai} (see also a review in Ref.~\cite{Panos}). In the
case of $a_{0}=-1$, solitons are {\it staggered} \cite{Cai}, with the $\pi$
phase difference between adjacent sites; thus, on the contrary to the
continuum nonlinear Schr\"{o}dinger (NLS) equation, the DNLS model supports
stable bright solitons for either sign (repulsion or attraction) of the
nonlinear interaction.

\begin{figure}[tbp]
\centerline{\psfig{figure=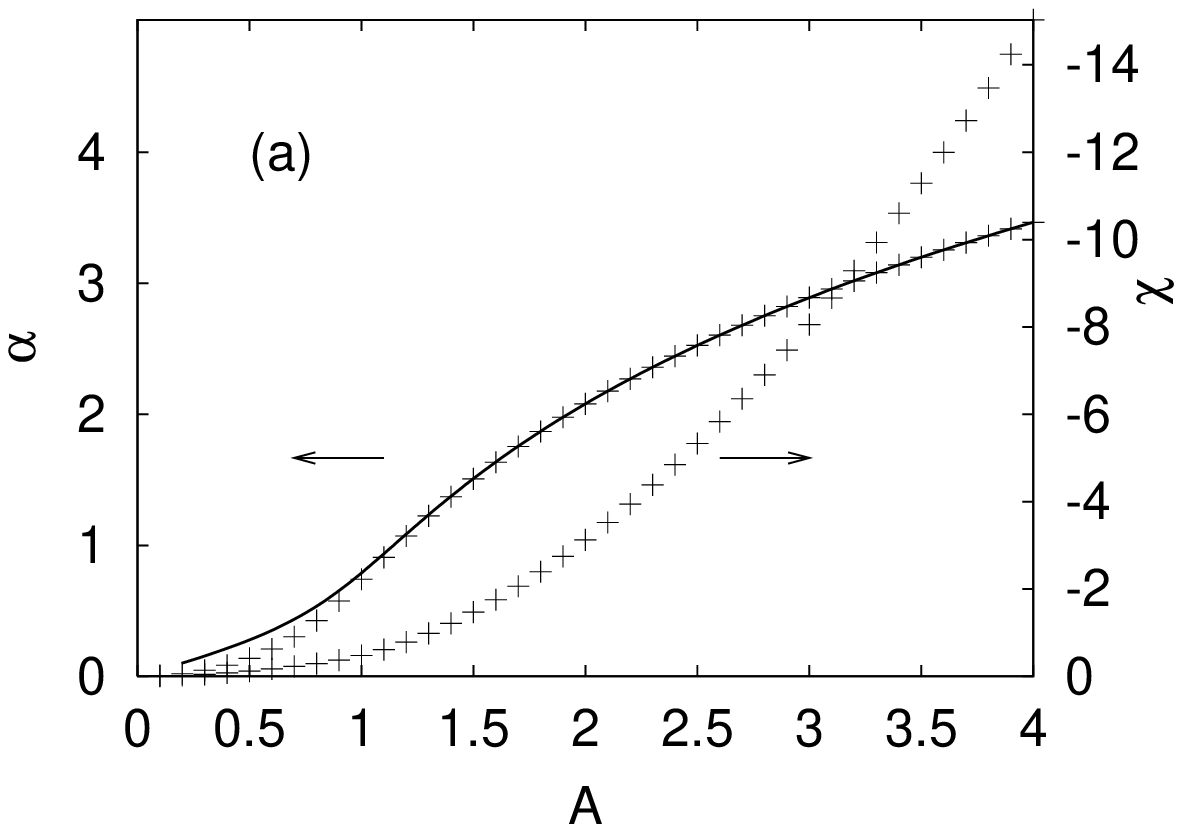,height=5.5cm}
  \psfig{figure=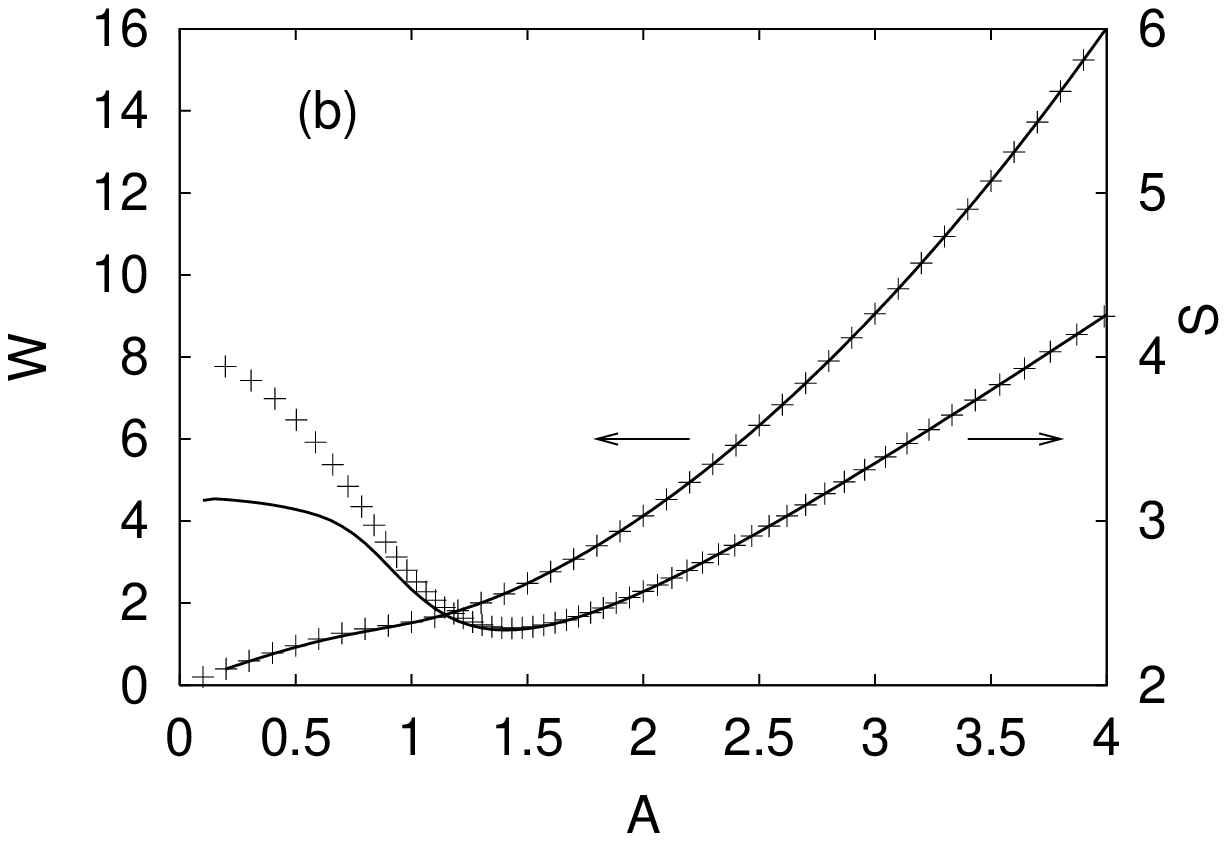,height=5.5cm}}
\caption{(a) The inverse width $\protect\alpha$\ (left axis) and the
frequency $\protect\chi$\ (right axis) of the soliton vs. its amplitude $A$
in the DNLS model without the time-modulation, $a_0=1$. (b) The norm $W$\
(left axis) and the soliton's area $S$\ (right axis) vs. $A$. Point symbols
represent data found from the numerical solution of the nonlinear eigenvalue
problem; the solid lines are the prediction of the analytical variational
approximation [see Eqs.(\ref{W}) and (\ref{alpha0})]. }
\label{f1-par}
\end{figure}

For convenience, here we briefly recapitulate basic properties of DNLS
solitons. Parameters of the discrete soliton (\ref{Soliton}), found
numerically from the nonlinear eigenvalue problem \cite{Scott}, are shown by
points in Fig.~\ref{f1-par}. The left axis in Fig.~\ref{f1-par}(a) pertains
to the inverse width $\alpha $, which was found by matching the soliton's
tail to the asymptotic expression $|u_{n}|=A\exp (-\alpha |n|)$, where $A$
is the soliton's amplitude [cf. Eq.~(\ref{Ansatz}) below]. The right axis in
Fig.~\ref{f1-par}(b) corresponds to the pulse's area, which we define as
$S=\sum_{n}{|}u_{n}{|}$. The dependencies for $a_{0}=-1$ have the same form,
with the only difference that $\chi $ is shifted so that $\chi \rightarrow
2-\chi $. Since $W$, $\chi $ and $\alpha $ are monotonic functions of $A$,
the stationary solution (\ref{Soliton}) is defined by fixing of any one of
these parameters.

\begin{figure}[tbp]
\centerline{\psfig{figure=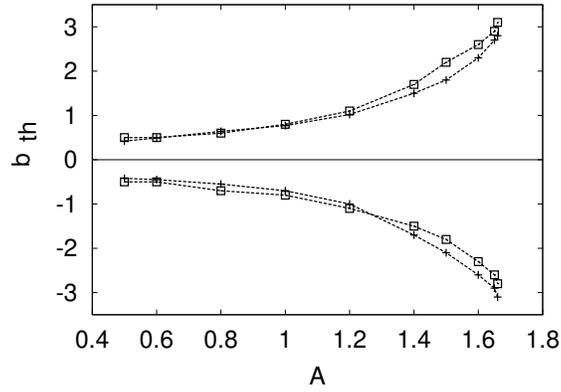,height=5.5cm}}
\caption{The critical value of the chirp added to the fundamental discrete
soliton, see Eq.~(\ref{IC1}), which splits the soliton into two separating
pulses, vs. the amplitude of the unperturbed fundamental soliton. Squares
(pluses) corresponds to $a_0= 1$\ ($a_0= -1$). }
\label{f2-chirp}
\end{figure}

Similarly to the case of the continuum NLS equation, the addition of chirp
to the soliton (\ref{Soliton}) ({\it chirp imprinting}) splits it into two
separating soliton-like pulses, if the chirp $b$ exceeds a critical
(threshold) value $b_{{\rm th}}$ (detailed consideration of a similar
problem in the continuum NLS equation was given in Ref.~\cite{Jim}). We
introduce the chirp by taking an initial condition as 
\begin{equation}
u_{n}(0)= Q_{n}\exp (ib|n|)  \label{IC1}
\end{equation}
(the value of $b$ is restricted to the interval $[-\pi ,\pi ]$). The
dependence of $b_{{\rm th}}$ on the amplitude $A$ of the unperturbed DNLS
soliton is presented in Fig.~\ref{f2-chirp}.

In fact, the curves shown in Fig. \ref{f2-chirp} diverge at sufficiently
large $A$. The meaning of this is that, if the soliton's amplitude $A$
exceeds the value $1.66$, the initial pulse with any amount of chirp gives
rise to a soliton centered at $n=0$, while other parts of the initial pulse
split off from it and move in opposite directions.

It is possible to understand the chirp-induced splitting of the pulse into
two in the following way. The original chirped pulse (\ref{IC1}) may be
regarded as a superposition of two pulses which carry the phase gradient of
opposite signs (cf. a similar model developed in the framework of the
continuum NLS equation in Ref.~\cite{Jim}). As is known, the velocity of an
isolated soliton is generated by its phase gradient. Since the two
constituents of the overall chirped pulse are originally close to each
other, the attraction between them is strong enough to keep them together.
However, the increase of $b$ leads to increase of the opposite
phase-gradient thrusts applied to the constituents, and finally to splitting
between them.

\section{The variational approximation and direct simulations in the case of
slow modulations}

\label{Sect:VA}

\subsection{The general formalism of the variational approximation}

The DNLS equation~(\ref{DNLS}) is derived from the Lagrangian, 
\begin{eqnarray}
L &=&\sum_{n=-\infty }^{\infty }{\frac{i}{2}}\left( u_{n}^{\ast }
\dot{u}_{n}-u_{n}\dot{u}_{n}^{\ast }\right) -  \nonumber \\
&&{\frac{1}{2}}|u_{n+1} -u_{n}|^{2}+\frac{1}{2}a(t)|u_{n}|^{4}\ .
\label{Ldef}
\end{eqnarray}
Following Ref.~\cite{Malom}, we base the variational approximation (VA) for
the soliton governed by Eq.~(\ref{DNLS}) on the following {\it ansatz}: 
\begin{equation}
u_{n}(t)=A\exp (i\phi +ib|n|-\alpha |n|)\ ,  \label{Ansatz}
\end{equation}
where $A,\phi ,b$, and $\alpha $ are real functions of time. Substituting
the ansatz (\ref{Ansatz}) into Eq.~(\ref{Ldef}), one can easily calculate
the corresponding {\it effective Lagrangian} in a form 
\begin{equation}
\frac{L}{W}=-\frac{1}{\sinh \left( 2\alpha \right) }\frac{db}{dt}+\frac{\cos
b}{\cosh \alpha }+\frac{1}{4}Wa(t)\frac{\sinh \alpha }{\cosh ^{3}\alpha }
\cosh \left( 2\alpha \right) \,,  \label{L}
\end{equation}
where 
\begin{equation}
W=A^{2}\coth \alpha ,  \label{W}
\end{equation}
is a dynamical invariant, which coincides with the total number of
particles, obtained by substitution of the ansatz (\ref{Ansatz}) into
Eq.~(\ref{norm}). We mention that a term in the full Lagrangian from
which it follows that $dW/dt=0$ contains the phase derivative
$\dot{\phi}$ [which gives the frequency $-\chi $ in the stationary
state, see Eq.~(\ref{Soliton})]. That term was dropped in the expression
(\ref{L}), as it does not contribute to other variational equations.
Finally, the variational equations for the soliton's chirp and inverse
width are
 \begin{eqnarray}
\frac{db}{dt} &=&2\left( \cos b\right) \frac{\sinh ^{3}\alpha }{\cosh \left(
2\alpha \right) }-\frac{1}{2}Wa(t)\left( \tanh ^{2}\alpha \right) 
\frac{2\cosh \left( 2\alpha \right) -1}{\cosh \left( 2\alpha \right) },
\label{db/dt} \\
\frac{d\alpha }{dt} &=&-\left( \sin b\right) \left( \sinh \alpha \right)
\tanh \left( 2\alpha \right) .  \label{da/dt}
\end{eqnarray}

\subsection{Revisiting the stationary model$\allowbreak $}

First, we dwell on the unperturbed case, with $a(t)={\rm const}\equiv a_{0}$
[cf. Eq.~(\ref{a(t)})]. In the case, all the points with $\alpha =0$ and $b=
{\rm const}$ are stationary solutions, i.e., fixed points (FPs). However,
they do not correspond to localized waves, therefore they are formal
solutions. Further, it is easy to see that Eqs.~(\ref{db/dt}) and (\ref
{da/dt}) give rise to nontrivial FPs with $b_{{\rm FP}}=0$ for $a_{0}=1$,
and $b_{{\rm FP}}=\pi $ for $a_{0}=-1$, the corresponding value 
$\alpha _{{\rm FP}}$ being defined by the equation \cite{Malom} 
\begin{equation}
\sinh \alpha _{{\rm FP}}=\frac{1}{4}W \left( 1+3\tanh ^{2}
\alpha _{{\rm FP}}\right) .  \label{alpha0}
\end{equation}
The parameters of the FP, which corresponds to stationary discrete soliton 
(\ref{Soliton}), are shown by solid lines in Fig.~\ref{f1-par}. As is seen,
the results of the VA are in good agreement with the exact numerical
solution of Eq.~(\ref{DNLS}). Deviation in $S$ (area of the pulse) indicates
that the VA is not applicable in the limit of small $A$. This is clear
because this limit corresponds to the continuum system, whose stationary
soliton solution differs from the ansatz (\ref{Ansatz}).

\begin{figure}[tbp]
\centerline{\psfig{figure=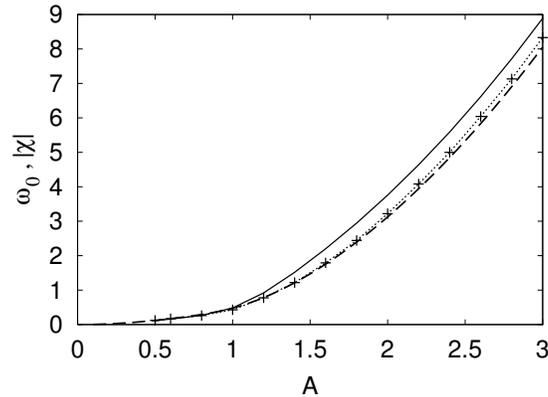,height=5.5cm}}
\caption{The frequency of small intrinsic oscillations of the discrete
solitons around the stationary configurations, in the case $a_0=1$, vs. the
soliton's amplitude $A$. The solid line shows the frequency 
$\protect\omega_0 $ as predicted, in the 
framework of the variational approximation, by
Eq.~(\ref{omega^2}). Points connected by the dotted line are values of the
forcing frequency which produce a resonant response in numerical simulations
of Eq.~(\ref{DNLS}) with a small time-periodic forcing term added to it. For
comparison, the dashed line shows the soliton's internal frequency
$|\chi|$.}
\label{f3-res}
\end{figure}

Linearization of Eqs.~(\ref{db/dt}) and (\ref{da/dt}) around the FP yields a
squared frequency of small oscillations, 
\begin{eqnarray}
\omega _{0}^{2} &=&\frac{\sinh ^{3}(\alpha _{{\rm FP}})\cosh ^{2}(\alpha _{
{\rm FP}})}{\cosh ^{3}(2\alpha _{{\rm FP}})}  \nonumber \\
\ \ \ \ \ \  &&\times \left\{ 4\sinh (\alpha _{{\rm FP}})[\cosh (2\alpha _{
{\rm FP}})+2]\right.   \nonumber \\
\ \ \ \ \ \  &&\left. -\frac{W}{\cosh ^{4}(\alpha _{\rm FP})}\left[ 5\cosh
^{2}(2\alpha _{{\rm FP}})-2\cosh (2\alpha _{{\rm FP}})-1\right] \right\} \;.
\label{omega^2}
\end{eqnarray}
Using Eq.~(\ref{alpha0}), one can show that $\omega _{0}^{2}$ given by
Eq.~(\ref{omega^2}) is always positive, i.e., VA does not predict any
(artificial) instability. The dependence of $\omega _{0}$ on $A$, obtained
from Eq.~(\ref{omega^2}), is shown by a solid line in Fig.~\ref{f3-res}. In
the same figure, crosses show resonant values of the frequency found from
numerical simulations of Eq.~(\ref{DNLS}) with a small coefficient $a_{1}$
in front of the variable part of the nonlinearity coefficient, see Eq.~(\ref
{a(t)}). In the simulations, the forcing frequency $\omega $ was varied at
the fixed small $a_{1}$, with the purpose to identify a value that generated
strongest resonant response. The relative difference between the thus found
resonance frequency and the value predicted by Eq.~(\ref{omega^2}) is about
$0.1$, and the overall behavior of the curves is identical. It is worthy to
note that $\omega _{0}$ almost coincides with the soliton's frequency $%
\left| \chi \right| $. Thus, the results presented in Figs.~\ref{f1-par} and 
\ref{f3-res} justify the validity of the VA based on the ansatz 
(\ref{Ansatz}).

\begin{figure}[tbp]
\centerline{\psfig{figure=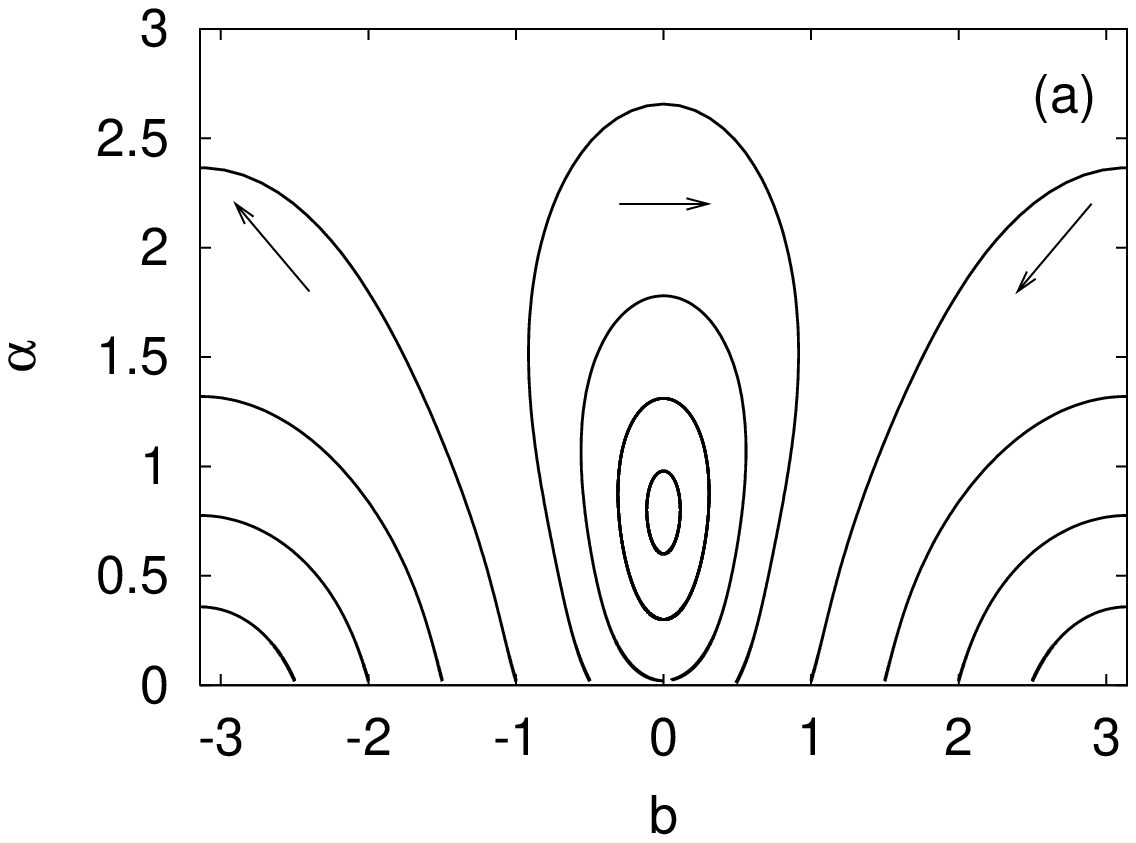,height=5.5cm}
  \psfig{figure=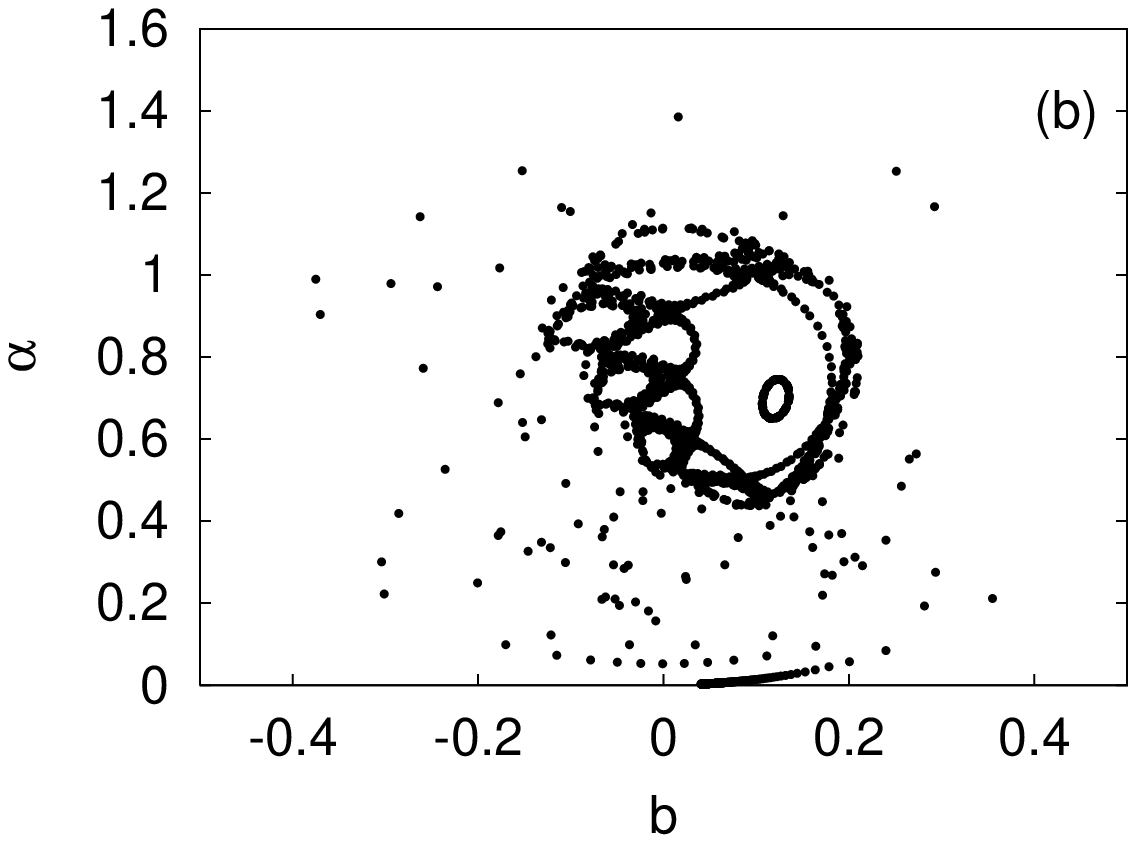,height=5.5cm}}
\caption{ (a) The phase plane of the dynamical system based on Eqs.~(\ref
{db/dt}) and (\ref{da/dt}), in the case of $a_0 = 1$, $a_1 = 0$, and $W=
1.5202$. Such a value of $W$ corresponds to a soliton with $A=1$. (b) An
example of chaotic dynamics for the periodically-modulated system at 
$W=1.5202$, $a_0 = 1$, $a_1=0.02766$ and $\protect\omega=0.481$.}
\label{f4-fp}
\end{figure}

The phase plane of Eqs.~(\ref{db/dt}) and (\ref{da/dt}) for $a_{0}=1$ and 
$a_{1}=0$ is shown in Fig.~\ref{f4-fp}(a), where arrows point out a direction
of motion along a trajectory. The phase plane for the case of $a_{0}=-1$ is
obtained by the shift $b\rightarrow \pi -b$, while that for the case 
$a_{0}=0 $ is obtained by setting $\alpha _{{\rm FP}}\rightarrow 0$. As it
follows from here, the stable FP, which corresponds to the discrete soliton,
exists for {\em either} sign of $a_{0}$, and vanishes if $a_{0}=0$. As it
also follows from Fig.~\ref{f4-fp}(a), the evolution initiated by the
initial condition with $\alpha (0)=\alpha _{{\rm FP}}$ and small $b(0)$
corresponds to oscillations near the FP. However, for large values of $|b(0)|
$, the asymptotic value of $\alpha (t)$ at $t\rightarrow \infty $ tends to
zero. This fact is in qualitative agreement with the above-mentioned result
that the addition of a chirp may destroy the soliton.

\subsection{The variational approximation for the nonstationary model}

We now proceed to the case of the ac-driven system, with $a_{1}\neq 0$.
If $a_{1}$ is small, strong response of the system to the time-periodic
modulation is expected when the modulation frequency $\omega $ is close
to the eigenfrequency $\omega _{0}$ of the internal oscillations of the
soliton in the unperturbed system, which is given by
Eq.~(\ref{omega^2}); in fact, the resonant response was already taken
into regard when collecting the data shown by crosses in Fig.
\ref{f3-res}. Moreover, the dynamics is expected to become chaotic, via
the resonance-overlapping mechanism, if the modulation amplitude $a_{1}$
exceeds some threshold value.

The Poincar\'{e} map illustrating a typical example of the chaotic
behavior, as found from the numerical solution of Eqs.~(\ref{db/dt}) and
(\ref{da/dt}), is presented in Fig.~\ref{f4-fp}(b). Shown are the
discrete trajectories initiated by sets of the initial conditions,
namely, the one with $(b_{1},\alpha _{1})=(0,0.789)$, that corresponds
to the stationary discrete soliton with $A=1$ in the unperturbed system
($a_{1}=0$), and $(b_{2},\alpha _{2})=(0.13,0.74)$. The modulation
frequency $\omega $ is close to the eigenfrequency of small oscillations
$\omega _{0}$. For the former initial condition, the point in the space
$(b,\alpha )$ is chaotically moving away from the unperturbed FP.
However, the chaotic evolution is a transient feature, as the point
eventually moves so that $\alpha (t)$ asymptotically tends to zero,
implying infinite broadening of the soliton.

As for the second set of initial conditions, a new FP is found in a
vicinity of the unperturbed one. This new FP predicts the existence of
quasi-stationary discrete FRM solitons in the case of the slow
modulation. Similar behavior near the corresponding stationary point is
observed for the case $a_0= -1$.

\subsection{Direct simulations}

We have performed systematic comparison of the predictions produced by the
VA against direct simulations of the full DNLS equation (\ref{DNLS}). The
simulations show that, generally speaking, VA correctly predicts only an
initial stage of the dynamics. The radiation of linear waves by a soliton,
which is ignored by the VA, gives rise to an effective dissipation, that
makes the resonance frequency different from $\omega _{0}$. Furthermore,
since $\omega _{0}$ depends on $W$, and the radiation loss results in
gradual decrease of $W$, the soliton decouples from the resonance.
In principle, VA might be made more accurate by adding a radiation mode
(``tail'') to the ansatz, cf. the analysis developed in Ref.~\cite{Smith}
for the soliton in the continuum NLS equation (see also the review \cite
{Progress}), but we do not aim to develop such an involved generalization of
the VA in the present work. In any case, a conclusion is that the dynamics
of the discrete soliton, as found from direct numerical simulation of 
Eq.~(\ref{DNLS}) for $a_{1}\lesssim 0.05$, is close to that predicted by the
variational equations (\ref{db/dt}) and (\ref{da/dt}). Namely, oscillations
of the soliton's parameters are regular for very small modulations, and
become chaotic when $a_{1}$ exceed a threshold, see below.

\begin{figure}[tbp]
\centerline{\psfig{figure=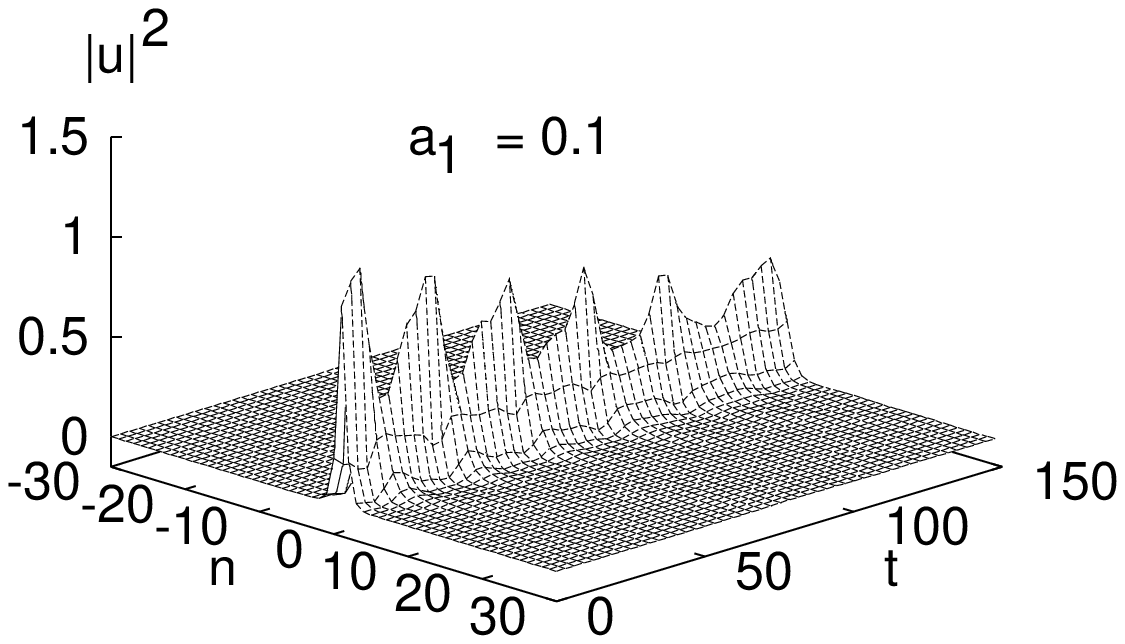,height=6cm}
  \psfig{figure=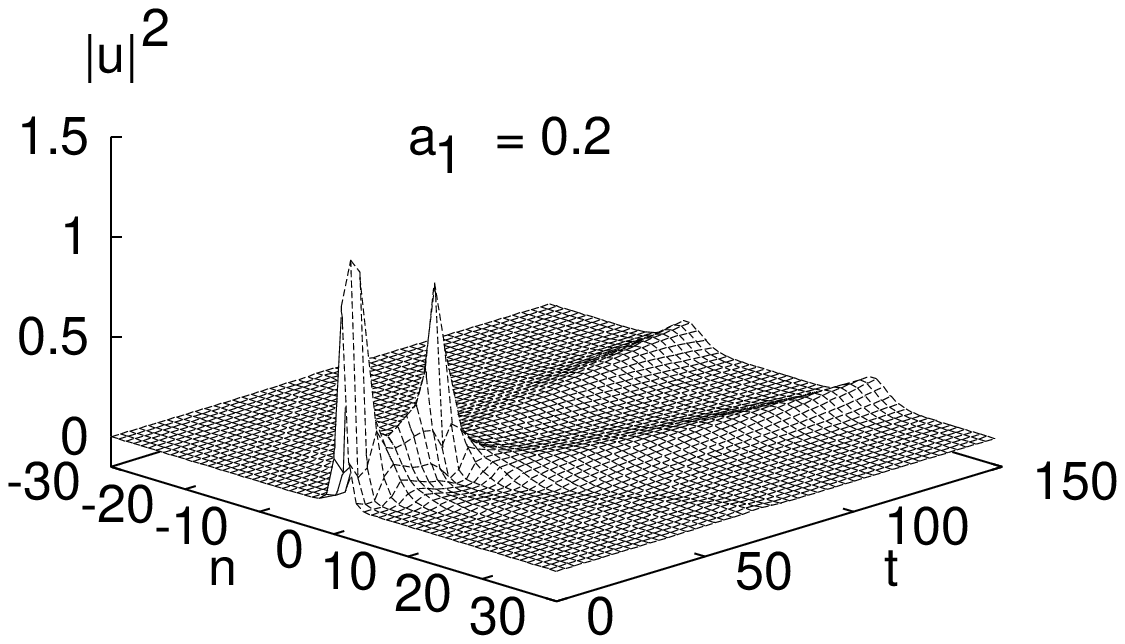,height=6cm}}
\centerline{\psfig{figure=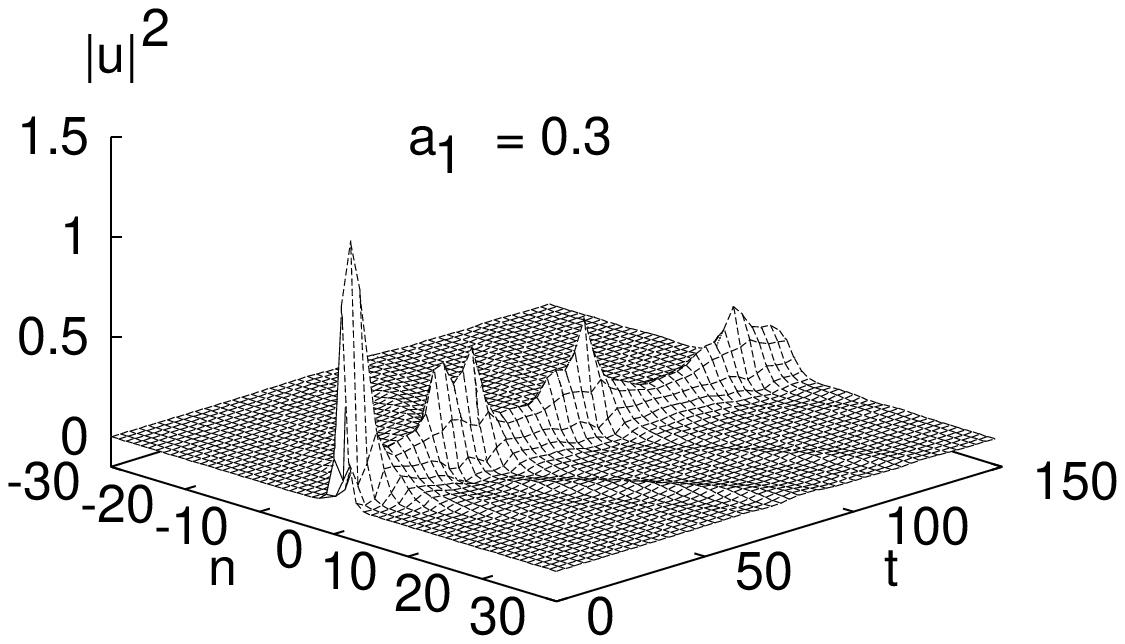,height=6cm}}
\caption{Evolution of a discrete soliton with the initial amplitude $A=1$ in
the periodically modulated system with $a_0=1, \protect\omega= 0.5$, and
different values of $a_1$.}
\label{f5-spl}
\end{figure}

Typical examples of the soliton dynamics with $\omega =0.5$ and
different values of the modulation amplitude $a_{1}$ are displayed in
Fig.~\ref{f5-spl}. An important observation, which is not predicted at
all by the single-soliton ansatz, is {\em splitting} of the pulse, which
is observed in Fig.~\ref{f5-spl}(b). Note that for other values of
$a_{1}$, in Figs.~\ref {f5-spl}(a) and \ref{f5-spl}(c), a stable soliton
is observed, centered at $n=0$, whose parameters oscillate because of
the modulation. Therefore, the splitting which occurs at $a_{1}\gtrsim
0.1$ is due to an interplay between the soliton itself, its intrinsic
eigenmodes, and the energy exchange with radiation modes (continuous
spectrum). It is noteworthy that the splitting is qualitatively similar
to that revealed by direct simulations of the continuum NLS equation
with a term accounting for periodic modulation of the linear dispersion
term [whose discrete counterpart is the finite-difference combination in
Eq.~(\ref{DNLS})], which was reported in Ref.~\cite{Ming}. A similar
phenomenon was also observed in the discrete model with the
finite-difference term subject to periodic modulation \cite{Pesch}.

\begin{figure}[tbp]
\centerline{\psfig{figure=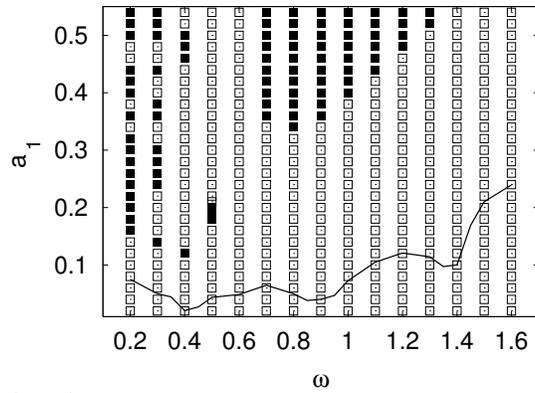,height=5.5cm}}
\caption{ The diagram in the plane $(\protect\omega, a_1)$, for the case 
$a_0= 1$. Open and solid rectangles correspond to stable and splitting
solitons, respectively. The initial soliton's amplitude is $A= 1$. The solid
line is the chaos-onset threshold as predicted by the variational equations. 
}
\label{f6-a1om}
\end{figure}

Results of the systematic numerical study of the splitting of the pulse
with the initial amplitude $A=1$ are summarized in Fig.~\ref{f6-a1om}.
In the simulations, absorbing boundary conditions, were used, the total
number of particles was $N \ge 200$, and the dimensionless simulation
time was, at least, $60\pi /\omega $. We classify as splitting cases
when at least two pulses emerge, moving in opposite directions, and no
pulse with an appreciable amplitude stays around $n=0$. For
$a_{1}\gtrsim 0.2$, the modulation results in generation of several
moving pulses. However, if a soliton with conspicuous amplitude is
eventually found around $n=0$, this case was classified as a ``stable
soliton''.

Figure~\ref{f6-a1om} also displays the dependence of a threshold
amplitude $a_{1}$, past which the initial state chaotically drifts to
$\alpha =0$, versus $\omega $ is also presented, as found from
simulations of Eqs.~(\ref {db/dt}) and (\ref{da/dt}). As is seen, the
splitting actually occurs far above the threshold in a region of the
developed dynamical chaos. The diagram for the case $a_0= -1$ looks
similar, but not exactly the same.


\section{The averaged equation for the case of rapid modulations}

In this section we consider the case of high-frequency modulations, with 
$\epsilon \equiv 1/\omega \ll 1$. Note that we do not require $a_{1}$ to be
small. In this case, it is natural to use the multiscale method \cite
{Kath,ACKMSU}. To this end, we introduce a set of time scales $\tau
=t/\epsilon $, $t_{k}=\epsilon ^{k}t$, where $k=0,1,2\dots $, and look for a
solution in the form 
\begin{equation}
u_{n}=U_{n} + \epsilon u_{n}^{(1)} + \epsilon^{2} u_{n}^{(2)}+\dots
\label{Expan}
\end{equation}
We substitute Eq.~(\ref{Expan}) into Eq.~(\ref{DNLS}) and collect terms at
the same order in $\epsilon $. Then, at order $\epsilon ^{0}$ we obtain 
\begin{equation}
i\frac{\partial U_{n}}{\partial t_{0}}+i\frac{\partial u_{n}^{(1)}}{\partial
\tau }+\frac{1}{2}\left( U_{n+1}+U_{n-1}-2U_{n}\right) +a(\tau
)|U_{n}|^{2}U_{n}=0\ ,  \label{eps0}
\end{equation}
where $a(\tau )\equiv a(t/\epsilon )$, and $U_{n}$ is a function of the slow
variables $t_{k}$. After averaging in the fast variable $\tau $, one has 
\begin{equation}
i\frac{\partial U_{n}}{\partial t_{0}}+\frac{1}{2}\left(
U_{n+1}+U_{n-1}-2U_{n}\right) +a_{0}|U_{n}|^{2}U_{n}=0\ ,  \label{dU0}
\end{equation}
where $a_{0}\equiv \left\langle a(\tau )\right\rangle $ standing for the
average value of the variable coefficient $a(\tau )$. Then the equation
for first correction $u_{n}^{(1)}$ takes the form
\[
i\frac{\partial u_{n}^{(1)}}{\partial \tau }=-[a(t)-a_{0}] \;
  |U_{n}|^{2}U_{n}\, ,
\]
a solution to which is 
\[
u_{n}^{(1)}=i\left( \mu _{1}-\left\langle \mu _{1}\right\rangle \right)
|U_{n}|^{2}U_{n}, 
\]
where $\mu _{1}\equiv \int_{0}^{\tau }[a(x)-a_{0}]dx$, and $\left\langle
...\right\rangle $ again stands for the average value.

At order $\epsilon^{1}$, we obtain $\partial U_{n}/\partial t_{1}=0$, and
\begin{eqnarray*}
&&u_{n}^{(2)}=(\mu _{2}-\left\langle \mu _{2}\right\rangle )\left[
|U_{n}|^{2}(U_{n+1}+U_{n-1})-\right. \\
\ \ \ \ &&\left. \frac{1}{2}U_{n}^{2}(U_{n+1}^{\ast }+U_{n-1}^{\ast })-\frac{
1}{2}|U_{n+1}|^{2}U_{n+1}-\frac{1}{2}|U_{n-1}|^{2}U_{n-1}\right] \\
\ \ \ \ &&-\frac{1}{2}[(\mu _{1}-\left\langle \mu _{1}\right\rangle
)^{2}-2M]|U_{n}|^{4}U_{n}\ ,
\end{eqnarray*}
where $\mu _{2}\equiv \int_{0}^{\tau }[\mu_{1}(x) - \langle \mu_{1}
\rangle] dx$, and $M=(\left\langle \mu_{1}^{2}\right\rangle
-\left\langle \mu _{1}\right\rangle ^{2})/2$. Finally, at order
$\epsilon ^{2}$ we find
\begin{eqnarray}
\frac{\partial U_{n}}{\partial t_{2}}
&=&iM[|U_{n+1}|^{2}(2|U_{n}|^{2}U_{n+1}+U_{n}^{2}U_{n+1}^{\ast })+
\nonumber\\
&&|U_{n-1}|^{2}(2|U_{n}|^{2}U_{n-1}+U_{n}^{2}U_{n-1}^{\ast })-  \nonumber \\
&&3|U_{n}|^{4}(U_{n+1}+U_{n-1})]+2iMa_{0}|U_{n}|^{6}U_{n}  \label{dU2}
\end{eqnarray}
Substituting Eqs.~(\ref{dU0}) and (\ref{dU2}) into the relation 
\[
\frac{\partial U_{n}}{\partial t}=\frac{\partial U_{n}}{\partial t_{0}}
+\epsilon \frac{\partial U_{n}}{\partial t_{1}}+\epsilon ^{2}\frac{\partial
U_{n}}{\partial t_{2}}+\dots \ , 
\]
one can derive the averaged equation, 
\begin{eqnarray}
&&i\dot{U}_{n}+\frac{1}{2}(U_{n+1}+U_{n-1}-2U_{n})+a_{0}|U_{n}|^{2}U_{n}=
\nonumber \\
\ \ \ \ \ \ &&-2Ma_{0}\epsilon ^{2}|U_{n}|^{6}U_{n}-M\epsilon ^{2}  \nonumber
\\
\ \ \ \ \ \times \ &&\left[
|U_{n+1}|^{2}(2|U_{n}|^{2}U_{n+1}+U_{n}^{2}U_{n+1}^{\ast })+\right. 
\nonumber \\
\ \ \ \ \ &&|U_{n-1}|^{2}(2|U_{n}|^{2}U_{n-1}+U_{n}^{2}U_{n-1}^{\ast })- 
\nonumber \\
\ \ \ \ \ &&\left. 3|U_{n}|^{4}U_{n+1}-3|U_{n}|^{4}U_{n-1}\right] \ ,
\label{av1}
\end{eqnarray}
where $M\equiv a_{1}^{2}/4$ for the case of the periodic modulation in
Eq.~(\ref{a(t)}).

Equation (\ref{av1}) is the higher-order DNLS equation produced by the
averaging procedure, which contains extra on-site and intersite (nonlocal)
nonlinearities. A change of variables $q_{n}\equiv U_{n}+\epsilon
^{2}M|U_{n}|^{4}U_{n}$ allows to rewrite Eq.~(\ref{av1}), retaining only
terms up to $O(\epsilon ^{2})$, in the following form 
\begin{eqnarray}
&&i\dot{q}_{n}+\frac{1}{2}(q_{n+1}+q_{n-1}-2q_{n})+a_{0}|q_{n}|^{2}q_{n}= 
\nonumber \\
\ \ \ &&\frac{1}{2}\epsilon ^{2}M\left[
3|q_{n}|^{4}(q_{n+1}+q_{n-1})+2|q_{n}|^{2}q_{n}^{2}(q_{n+1}^{\ast
}+q_{n-1}^{\ast })+\right.  \nonumber \\
\ \ \ \ \ \ &&\left. |q_{n+1}|^{4}q_{n+1}+|q_{n-1}|^{4}q_{n-1}\right] - 
\nonumber \\
\ \ \ &&\epsilon
^{2}M[|q_{n+1}|^{2}(2|q_{n}|^{2}q_{n+1}+q_{n}^{2}q_{n+1}^{\ast })+  \nonumber
\\
\ \ \ \ \ \ \ &&|q_{n-1}|^{2}(2|q_{n}|^{2}q_{n-1}+q_{n}^{2}q_{n-1}^{\ast
})]\ .  \label{qeq}
\end{eqnarray}
An advantage of the equation in the form (\ref{qeq}) is that it can be
derived from a Lagrangian, 
\begin{eqnarray}
&&L_{q}=L_{0}-\frac{1}{2}\epsilon ^{2}M\sum_{n=-\infty }^{\infty
}(|q_{n+1}|^{2}-|q_{n}|^{2})^{2}  \nonumber \\
\ \ \ \ \times &&(q_{n}^{\ast }q_{n+1}+q_{n}q_{n+1}^{\ast })\ ,  \label{Lq}
\end{eqnarray}
where $L_{0}$ is obtained from the underlying Lagrangian~(\ref{Ldef}) by the
substitution $u_{n}\rightarrow q_{n}$ and $a(t)\rightarrow a_{0}$. The
existence of the Lagrangian $L_{q}$ allows one to apply the variational
approximation (VA) like in Section~\ref{Sect:VA}.

\begin{figure}[tbp]
\centerline{\psfig{figure=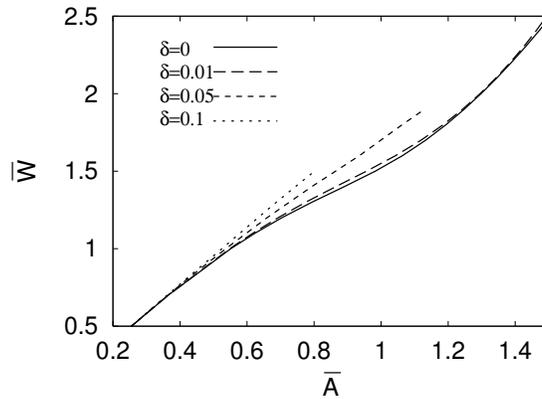,height=5.5cm}}
\caption{ The dependence of $\bar{W}$ vs. $\bar{A}$ of an average soliton
(dashed lines) is compared with that of the unperturbed DNLS equation (solid
line), $a_0=1$. }
\label{f7-aver}
\end{figure}

For the application of VA, we take the ansatz for $q_{n}$ in the form 
\begin{equation}
q_{n}=B\exp (i\psi +ic|n|-\beta |n|),  \label{Ansq}
\end{equation}
cf. Eq.~(\ref{Ansatz}). Substituting Eq.~(\ref{Ansq}) into Eq.~(\ref{Lq}),
we calculate the effective Lagrangian, 
\[
L_{q}=L_{0}-4\epsilon ^{2}MW_{q}^{3}\cos (c)\frac{\sinh ^{2}(\beta )\tanh
^{3}(\beta )}{\sinh (3\beta )}\ . 
\]
Here $L_{0}$ is the same expression as in Eq.~(\ref{L}), with a change
$b\rightarrow c$, $\alpha \rightarrow \beta $, $W\rightarrow
W_{q}=B^{2}{\rm coth}(\beta)$, and $a(t)\rightarrow a_{0}$. Now one can
deduce a dynamical system for the variable $c$ and $\beta $ similar to
Eqs.~(\ref{db/dt}) and (\ref{da/dt}). The fixed point $(\beta
_{{\rm FP}},0)$ for $a_0=1$, or $(\beta _{{\rm FP}},\pi)$ for $a_0=-1$,
of this system represents a FRM soliton in the case of rapid
modulations, where $\beta _{ {\rm FP}}$ is to be found from the equation
 \begin{eqnarray}
&&\sinh (\beta _{{\rm FP}})-\frac{W_{q}}{4} [1+3\tanh ^{2}
  (\beta _{{\rm FP}})]+ \nonumber \\
&& 4 {\rm sign}(a_0) \epsilon ^{2}MW_{q}^{2}\sinh (\beta _{{\rm FP}})\;
  \tanh ^{2} (\beta _{{\rm FP}})  \nonumber \\
&&\times {\frac{\lbrack 10+15\cosh (2\beta _{{\rm FP}})-\cosh
  (4\; \beta _{{\rm FP}})]}{[1+2\cosh (2\beta _{{\rm FP}})]^{2}}}=0\; .
\label{averFP}
\end{eqnarray}
The norm $\bar{W}$ and amplitude $\bar{A}$ of the field $U_{n}$ in the
averaged soliton are related to those of the field $q_{n}$ as 
\begin{eqnarray}
\bar{W} &\approx &W_{q}[1-2\epsilon ^{2}MW_{q}^{2}\left( \tanh ^{3}\beta
\right) \coth (3\beta )]\;,  \nonumber \\
\bar{A} &\approx &A_{q}(1-\epsilon ^{2}MA_{q}^{4})\,.  \label{relat}
\end{eqnarray}
The dependence $\bar{W}(\bar{A})$ found from Eqs.~(\ref{averFP}) and (\ref
{relat}) at different values of $\delta \equiv a_{1}^{2}/(4\omega ^{2})$ is
displayed in Fig.~\ref{f7-aver}. Different curves in the figure terminate at
finite values of $\bar{A}$ because the relation (\ref{relat}), as well as
the change of variables $U_{n}\rightarrow q_{n}$, are not valid outside the
corresponding intervals. As it is suggested by Fig.~\ref{f7-aver}, one can
effectively control the soliton by an appropriate choice of the modulation
parameters. Increase of the total number of particles in the averaged
soliton, as compared to that in the unperturbed soliton with the same
amplitude, is clearly seen in Fig.~\ref{f7-aver}.

\section{Conclusions}

We have studied the dynamics of an array of Bose-Einstein condensates with
the time-dependent scattering length. Applying the variational
approximation, the frequency $\omega _{0}$ of small intrinsic oscillations
of the soliton was predicted. The possibility of chaotic dynamics in the
near-resonance case, when the driving frequency $\omega $ is close to 
$\omega _{0}$, was shown. Direct simulations have demonstrated that the
modulations of sufficient strength may result in splitting of the soliton.
Results of the simulations were summarized in the form of the diagram which
shows the splitting regions in the $\left( \omega ,a_{1}\right) $ plane. The
existence of stable Feshbach-resonance-managed discrete matter-wave solitons
was demonstrated in the cases of both slow and rapid modulation of the
nonlinearity coefficient. In the latter case, the soliton dynamics reduces
to the generalized DNLS equation, which involves additional on-site and
inter-site nonlinearities. By making use of this equation, properties of the
averaged soliton were predicted. In particular, increase of the total number
of atoms in this soliton in comparison with the ordinary discrete soliton of
the same amplitude was shown.

The {\it chirp imprinting} discussed in Sect.~\ref{Sect:Gen} can be an
effective tool, similar to the phase-engineering method \cite{Phase}, for
manipulating the condensate's wave function. Pulse splitting induced by the
chirp imprinting, or otherwise by the application of Feshbach-resonance
modulation, can be used as a source of coherent pulse pairs in an atomic
Mach-Zehnder interferometer \cite{Poet}.

\section*{Acknowledgement}

B. A. M. appreciates hospitality of Instituto de Fisica Teorica at UNESP
(Sao Paulo, Brazil). F. Kh. A. is grateful to FAPESP for the partial support
of this work.

\newpage

\end{document}